\documentclass{aa}
\usepackage{graphics}
\usepackage{psfig}

\begin{document}

\thesaurus{07(07.16.1; 17.16.2;Saturn)} 

\title{Detection of arcs in Saturn's F ring during the 1995 Sun ring-plane crossing}
   \titlerunning{Observations of Sun ring-plane crossing}

\author{S.~Charnoz$^{1}$, A.~Brahic$^{1}$, C.~Ferrari$^{1}$, I.~Grenier$^{1}$, F.~Roddier$^{2}$, P.~Th\'{e}bault$^{3}$
}
	\authorrunning{S.~Charnoz et al.}

\offprints{S. Charnoz}
\mail{charnoz@cea.fr}

\institute{Equipe Gamma-Gravitation,
Service d'Astrophysique, CEA/Saclay, Orme des Merisiers, 91191 Gif sur Yvette cedex, France
\and
Institute for Astronomy, University of Hawai, 2680 Woodlawn Drive, Honolulu, Hawaii 96822
\and
DESPA, Observatoire de Paris, 92195 Meudon Cedex Principal, France
}

   \date{Received date / Accepted date}
   \maketitle

\begin{abstract}
Observations of the November 1995 Sun crossing of the Saturn's ring-plane made with
the 3.6m CFH telescope, using the UHAO adaptive optics system, are presented here.
We report the detection of four arcs located in the vicinity of the F ring.
They can be seen one day later in HST images. The combination of both data sets gives
accurate determinations of their orbits. 
Semi-major axes range from 140\,020 km to 140\,080 km, with a mean of $140\,060 \pm 60$ km.
This is about 150 km smaller than previous estimates of the F ring radius 
from Voyager 1 and 2 data, but close to the orbit of another arc observed at the same epoch
in HST images.
\end{abstract}

\section{Introduction}
The Sun's crossing of Saturn's ring-plane is a rare opportunity to observe the 
unlit face of the rings. Results of observations made at the 3.6m CFH Telescope are reported here.
Thanks to the use of the UHAO adaptive optics system (Roddier et al. \cite{roddier_2000}), 
high resolution ground-based images were obtained ($\sim$0.15\arcsec\,in August 1995 and 
$\sim$0.25\arcsec\,in November 1995). The Sun's crossing lasted from November 17$^{th}$ to 21$^{st}$ 1995. It was accompanied
 by three Earth ring plane crossings (1995 May 22$^{nd}$, August 10$^{th}$, and 1996 February 11$^{th}$). Several teams observed these events (Nicholson et al. \cite{nicholson_96}, hereafter N96; Bosh
 \& Rivkin \cite{bosh}; Sicardy et al. \cite{sicardy}; Roddier et al. \cite{roddier_96a}; 
Ferrari et al. \cite{ferrari}).\\
The unusual viewing of the rings during these events provides a rare opportunity to detect faint 
structures  such as small satellites, clumps, or arcs, 
that are usually lost in the glare of the bright main rings. In this paper, we focus 
our attention on the azimuthal structure of the F ring. Located at the Roche limit of Saturn and 
bounded by two shepherding satellites Pandora and Prometheus,
the F ring is very intriguing. The Voyager images revealed its complex radial and azimuthal
structure including  multiple strands, clumps and braids. Interactions with Saturn's satellites might
explain some features, but they are not completely understood.\\
During the 1995 ring crossing observing campaign, the F ring  appeared populated with 
numerous features, either point-like or longitudinally extended. The observation of two point-like 
objects (S/1995 S1 and S/1995 S3) was reported by Bosh \& Rivkin (\cite{bosh}) during the May crossing.
During the August crossing, our team discovered at least 6 new 
features with semi-major axes compatible with the
 F ring (Roddier et al. \cite{roddier_96a}; Ferrari et al. \cite{ferrari}). At least one of them was 
identified as an arc-like object. During the same period three objects 
(S/1995 S5, S/19995 S6 and S/1995 S7) were detected by N96. 
Finally, during the November crossing, two wide arcs ($7\degr$ and 10\degr-long) were observed
by N96. The brightest was also seen by Poulet et al. (\cite{poulet}, hereafter P2000). The most striking
aspect of those discoveries is that no evident correlation can be found between features observed in 
May, August and November (N96, Bosh \& Rivkin 1996, P2000). 

In the present paper, we report the detection of 4 arcs in the region of the F ring on November 20$^{th}$. 
The same features can be seen in some HST images taken 24 to 36 hours later. These images have been already 
presented in N96. By combining both sets of data, accurate orbital solutions are derived. The data processing of 
CFHT and HST images is presented in the first section. Results are presented in the second section and 
discussed in the final section.

\section{Description of data set and processing}

\begin{figure*}
\centering
 \resizebox{12cm}{!}{\includegraphics{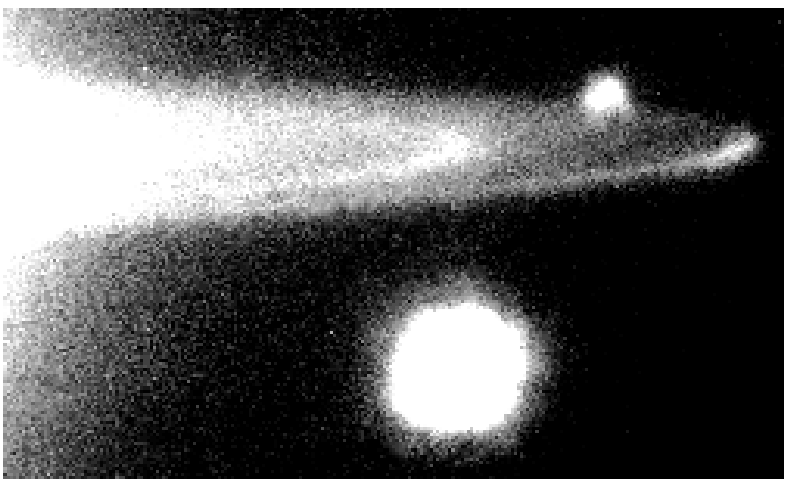}}
 \hfill
 \caption{ Image of the West ansa in the K band on 1995 November 21 at 5h11 UTC, obtained at CFHT 
with the UHAO system.  The F ring and the Cassini 
division appear brightly. Janus (upper right) and Tethys (bottom center) are visible in this image. 
The north pole of Saturn is
oriented toward the top of the image. This image is slightly satured as can be seen 
from the Cassini division and the F ring.}
\label{Fig.1}
\end{figure*}

\subsection{Description of data}
Images were taken at the 3.6m CFH Telescope during four nights from 1995 
November 17$^{th}$ to 20$^{th}$. Three near-infrared filters (J,H and K), which correspond 
to the absorption bands of methane were used to significantly reduce the strong scattered light of Saturn. A K band image is presented in Fig~\ref{Fig.1}. The best observing conditions were obtained during the last night, 
the main rings being 
darkest. During the nights of November 17$^{th}$,18$^{th}$19$^{th}$ the stronger brightness of the
main rings was not favorable for the detection of dim objects. Consequently only images of November 20$^{th}$ were considered. 

At this time the Sun was about 0.01\degr\, 
below the rings, and the Earth was at 2.67\degr\,above the ring plane, looking at the unlit face. 
160 images of the west ansa were obtained from 5h to 8h45 UT. At this time, there 
was no satellite on the east ansa bright enough for the adaptive optics tracking
system to work properly. 
In consequence, the CFHT data set contains only images of the west ansa. 
The resolution has been checked on every image 
by extracting the point spread function (PSF) of satellites Tethys, Enceladus, Dione and Mimas. 
The full width at half maximum of the PSF is on average 0.25\arcsec\,
(1670 km at Saturn) on the whole data set, ranging from 0.19\arcsec to 0.33\arcsec. The size
of one pixel in arcsecond is determined by comparing the angular separation of all pair of major satellites 
(given by the ephemeris) with their separation in pixels as measured on the plates. The result
is robust : one pixel subtends 0.035\arcsec\, ($\sim$ 233 km at Saturn). 
The exposure times ranged from 15 s to 30 s.\\
In order to make a comparative study of two complementary data sets, we have also used the HST images 
presented in N96, retrieved from the FTP site of the Space Telescope Science Institute.
They consist of twenty Planetary Camera images of both ansae (10 images for each ansa),
taken with filters centered at 890 nm, between 10h UT and 18h UT on November 21$^{st}$, at the end of the
 Sun's crossing.
The resolution is 0.1\arcsec\,(668 km at Saturn). The scale of the plates is taken from Holtzman et al. 
(\cite{holtzman}): one pixel subtends 0.045\arcsec\, ($\sim$304 km at 
Saturn). Exposure times range from
300s to 500s. As a consequence, satellites appear smeared due to their keplerian motion.

\subsection{Data processing}
Our data processing has been designed for the detection of faint objects in the F ring. As a consequence we
have not made a photometric study; indeed many images with different exposure times and 
different filters are combined indiscriminately.
A longitudinal profile of the F ring was built to reveal new objects. To this purpose the
following data processing steps were applied to both CFHT and HST images : 
(i)astrometry, (ii)extraction of the F ring profiles, (iii) subtraction of the F ring background and the correction 
of geometry and illumination effects.

\subsubsection{Astrometry}
In CFHT images, the center of Saturn and orientation of the planet's North pole axis is determined 
by a least square fit to the positions of satellites Tethys, Enceladus, Mimas and Dione 
(using the ephemeris of the Bureau Des Longitudes in Paris). 
However Mimas and Tethys could not be used in one third of images in which they are observed too
close to each other, not suitable for an accurate determination of their center. In these images
the point of maximum elongation of the Cassini division and of the F ring were used to better constrain the
orientation of Saturn's equatorial axis.
The residual on satellites position measurements is 0.09\arcsec 
($\sim$2.6 pixels or 600 km), and reflects a combination of pointing and ephemeris errors.\\

In HST images, the astrometry has not been done using satellites because of their elongated appearance. 
The sharp inner edge of the Cassini Division was used instead as a reference. 
Located at 117\,577 km from the center of Saturn, it is known to be slightly eccentric due to the 
close 2:1 resonance with Mimas. The resulting radial excursion (2ae) is  only 150 km, 
representing  0.5 pixel, or 0.02 \arcsec (Porco et al. \cite{porco}). 
The center of Saturn as well as the orientation of the axes are determined by fitting an ellipse with 
twenty points picked by eye on the reference edge. 
Corrections of the geometrical distortion for WF/PC2 camera (WF/PC2 instrument handbook version 5.0, provided by 
the STSCI) is included only when
new objects are measured, for the need of their orbit determination. The astrometric error is determined from the residuals of the Cassini division fit. It is 0.035\arcsec ($\sim$0.8 pixel or 240 km at Saturn).
The correction for geometrical distortion is  about 1 pixel, representing less than $0.6\degr$ 
on the F ring longitudinal profile, not detectable in Fig~\ref{Fig.2} and Fig~\ref{Fig.3}.

\subsubsection{Extraction of the F ring profile}

From -55\degr\, to +55\degr\, around maximum elongation, pixel counts are summed perpendicularly 
to the equator axis, in a box centered on the F ring (defined as a circle of radius 140\,200 km 
and projected onto the ring plane) with the same width as the PSF. For each image, 
an intensity profile of the F ring is obtained as a function of longitude, measured from the 
point of maximum elongation. 
The extremity of the ansa
(within 5 pixels around maximum elongation) is not considered because the scans become parallel to the ring 
in this region. The scattered light of 
Saturn is estimated in the neighborhood of the integration box, outside the F ring, and systematically subtracted. 
Profiles are then corrected for effects of the different filters and for different exposure times: each profile 
is multiplied by a normalization factor 
in order to keep at the same median level of brightness a reference zone of the F ring, chosen 
to be between 0\degr\, and 45\degr\, before maximum elongation. Because of the low opening angle 
of the rings system, 
the reference zone is slightly contaminated by the A ring beyond 25\degr\ from maximum elongation. 
However we have checked that all profiles do not present systematic deviations in the reference zone after this
operation.

\subsubsection{ Subtraction of the F ring background and correction of illumination effects}
A synthetic median profile (SMP) of the F ring is built by taking the median value of 
all profiles at each longitude relative to elongation. The SMP is a template of the F ring. It is 
subtracted from all profiles in order to reveal local brightness variations. \\
The apparent brightness of the F ring changes along its path around the ansa due to
 geometrical effects (the length of a ring segment included in one pixel
 depends on longitude) and probably to shadowing by the A ring (N96). Indeed the F ring brightness 
decreases abruptly after maximum elongation on the west ansa, as noticed in N96. 
To correct for these effects and  maintain embedded features at a constant level 
of brightness, all profiles 
are divided by the SMP. Taking into account the subtraction of the F ring background,
 all profiles are processed according to the formula: Final Profile $=(profile-SMP)/SMP$. 
Dividing by the SMP increases noise in dim regions. This method preserves the 
brightness of objects within 20\% before and after their passage through the maximum 
elongation (as observed on arcs B,C and on the 10\degr\ arc of N96, see below) and allows tracking of objects.

\subsubsection{Construction of the longitudinal F ring profile}
\begin{figure*}
\centering
 \resizebox{12cm}{!}{\includegraphics{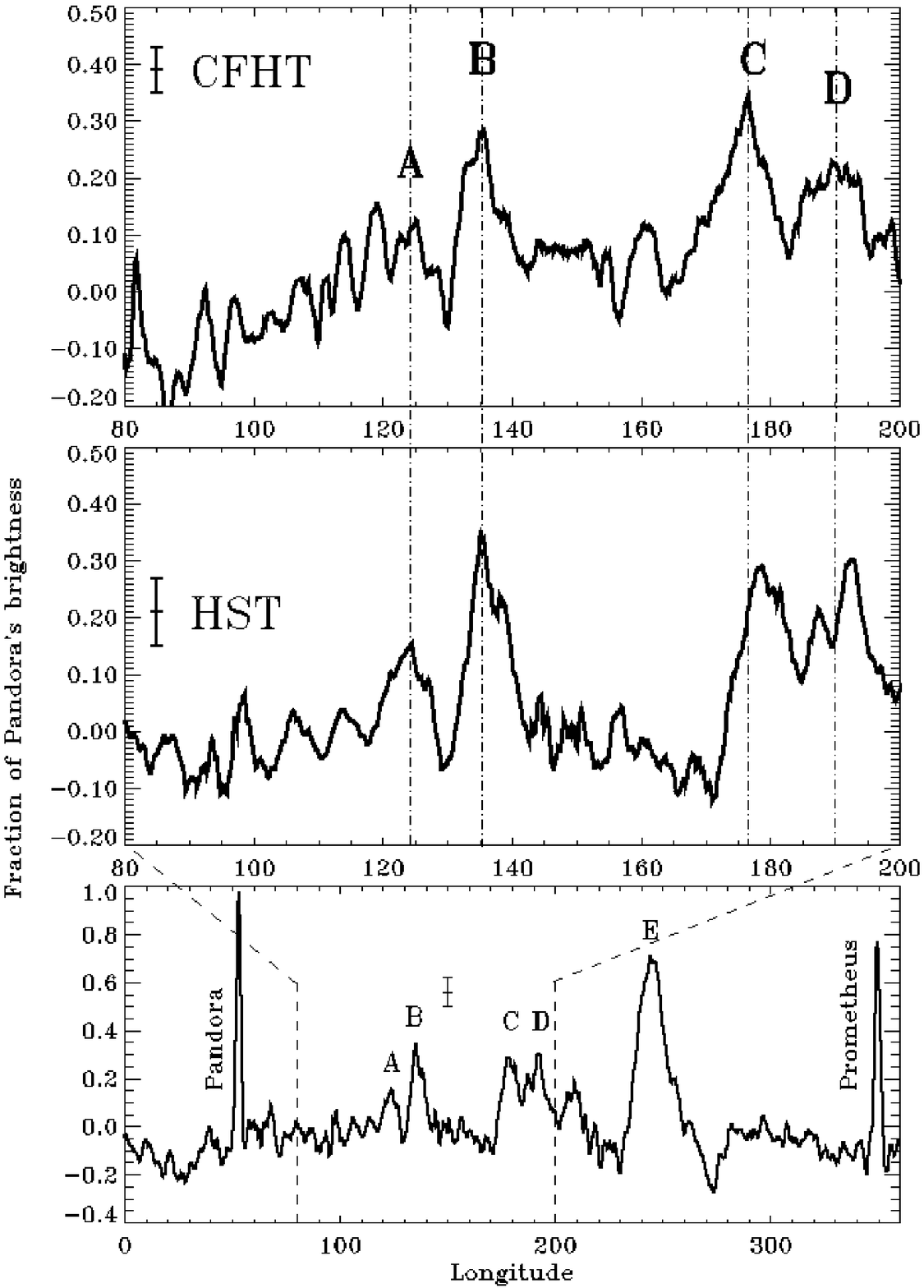}}
 \hfill
 \caption{Longitudinal brightness profiles extracted from 160 CFHT images of the west ansa and 20 HST images of both ansae (2.a and 2.b respectively) in the common longitudinal range.
Fig 1.c is the full HST profile with complete azimuthal coverage, obtained with
our data processing. Error bars are $\pm$1$\sigma$. 
Negative values mean that brightness is locally lower than the background
of the ring. All profiles are precessed to the standard epoch 1995 November 21.5 TDT at Saturn. 
The B and E arc are refered by N96 as the ``7\degr\ arc'' and the ``10\degr\ arc'' respectively. 
The orbit of the E arc has been determined by P2000.
}
\label{Fig.2}
\end{figure*}

Each (final) profile is precessed to the standard epoch of November 21.5 TDT (terrestrial dynamic time) 
at Saturn, with a mean orbital motion of
 $582.05\degr$/day (N96). This allows direct comparison with N96 results. Finally a mean longitudinal profile 
is computed by averaging all precessed profiles. \\
The resulting CFHT and HST profiles are presented in Fig.~\ref{Fig.2}.  The derived HST full azimuthal profile
(see  Fig.~\ref{Fig.2}.c) is in close agreement with the one published in N96, in terms of the positions of 
objects and 
relative brightnesses, although our profile may appear somewhat noisier than the one published in N96.
In this profile, the correction for the distortion of the WF/PC2 camera has not been considered, resulting
in an azimuthal error less than $\sim$0.6\degr. The vertical axis is scaled such that the maximum intensity of Pandora, 
as observed in the processed profiles, equals 1.
\section{Results}

\begin{figure*}
\centering
 \resizebox{12cm}{!}{\includegraphics{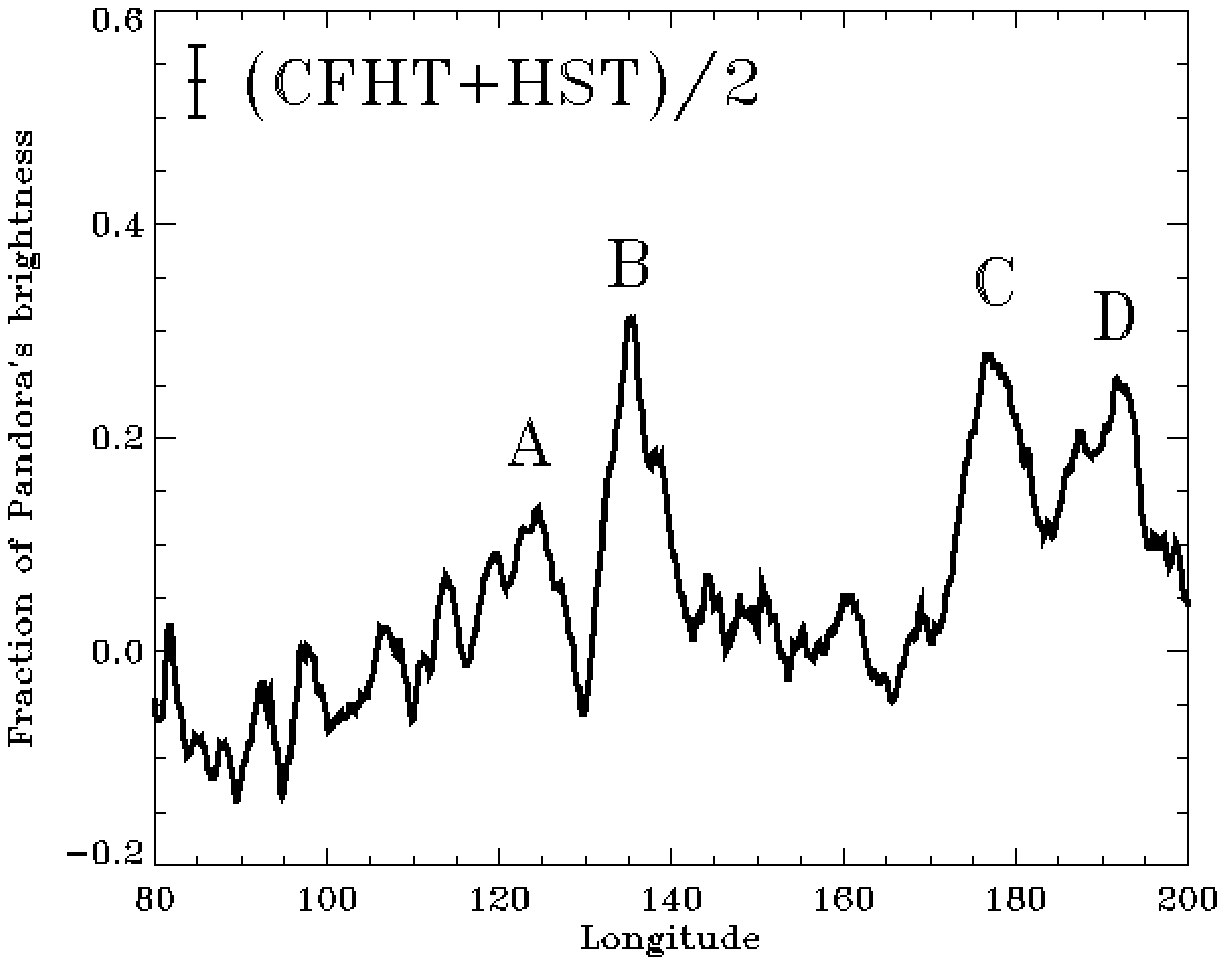}}
 \hfill
 \caption{Mean of CFHT and HST profiles presented in Fig.~\ref{Fig.2}.a. and Fig.~\ref{Fig.2}.b. in the 
common azimuthal region at the standard epoch. Error bar is
$\pm$1$\sigma$.}
\label{Fig.3}
\end{figure*}

In this section the F ring azimuthal profile obtained from CFHT observations is described 
and compared with the HST profile. Orbital solutions for four detected structures are then calculated.
 
\subsection{ F ring longitudinal profile and new objects}

\subsubsection{CFHT images}
The complete azimuthal coverage of the data set ranges between 50\degr\, to 230\degr. 
But the region extending from 50\degr\, to 80\degr has been covered by only a few suitable images ($\sim$10) and 
is deeply embedded in the very strong scattered light of Saturn's globe. 
Apart from Pandora appearing at 62\degr, nothing is detected below 80\degr. 
Janus contaminates CFHT profiles around 210\degr\, longitude. The region of interest 
then extends between 80\degr\, and 200\degr\ (Fig ~\ref{Fig.2}.a). In this range, the average error bar 
($1\sigma$) is about 4\% of Pandora's brightness.\\
Two structures (B and C) are detected with more than $5\sigma$ confidence level, 
with mean central longitudes of 135\degr and 177\degr respectively. The B object is the ``7\degr\ arc''
in N96. Their 
maximum brightness is 30\% to 40\% of Pandora's. Compared to the $3\degr$-azimuthal 
extension of a point-like source due to the PSF (as observed for Pandora), it is 
clear that B and C are azimuthally extended objects. The B object has a FWHM of $\sim$ 7\degr. 
It is observed before (from 5h10 to 6h30 UT) and after its passage through maximum 
elongation (from 7h40 to 8h45 UT). Feature C has a FWHM of 10\degr. It is detected during 20 minutes before
maximum elongation (from 5h10 to 5h30) and during 80 mn after maximum elongation (from 5h50 to 7h10 UT).
A few other bumps also seem to appear with lower confidence level 
($\sim$\,2-3$\sigma$) below 120\degr\ and at 124\degr, 163\degr and 190\degr. 
Only two of them may be tracked on many images with a keplerian motion. 
The A object, located at 124\degr, is nearly marginal here
and seems to extend from 117\degr\,to 129\degr\,and has an intensity about 15\% of Pandora's. 
It is detected from 5h50 to 6h40 UT ($\sim$ 30 images). 
The D object (located at 190\degr) may be followed from 5h40 to 6h20 UT.
Its azimuthal shape is uncertain because of numerous nearby artifacts on the images where it is detected.

\subsubsection{Comparison with HST data}
\label{HST}
Comparison of CFHT and HST profiles indicates that at least structures B and C lasted over a
24 to 36 hours time interval. Indeed both profiles are in good agreement 
within $1\sigma$ (see Fig.~\ref{Fig.2}.a and Fig.~\ref{Fig.2}.b).
Because of the small number of images, the noise varies significantly 
with longitude in the HST profile, depending on the number of images covering each 
region. It is on average $\sim$ 6\% of Pandora's brightness. The azimuthal extension of 
a point-like object is also 3\degr (measured on Pandora and Prometheus), but here it 
is mainly due to smearing. Objects B and C are present in both profiles. 
The shape and position of object B fits particularly well. 
C appears slightly narrower in the HST profile, but both are compatible within  $1\sigma$ in intensity. 
Its FWHM is about 9\degr\, in the HST profile. \\
A better signal to noise ratio is simply obtained by averaging CFHT and HST profiles 
(see Fig.~\ref{Fig.3}). The new error bar at $1\sigma$ is $3.3\%\,$of Pandora's 
brightness,which improves by factors 1.2 and 1.8 upon the signal to noise ratio of CFHT and 
HST profiles respectively. In the averaged profile the confidence level of objects B and C 
increases to $9\sigma$ and  $8\sigma$ respectively. Objects A and D, that were marginal
in the CFHT profile, appear now at $5.5\sigma$ and $5\sigma$ respectively, showing these
are real structures. Whereas the shape of A is in good agreement between CFHT and HST profiles, 
the shape of D is more peaked in the HST profile, and is centered at 192\degr. 
Features below 120\degr and at 164\degr are comparable to the error bar in the averaged profile. The 
simultaneous presence of structures A,B,C,D with same positions and similar shapes 
in both CFHT and HST data confirms that these are real and extended objects. 
They are designated as ``arcs'' hereafter.

\subsection{Orbital fit}

\begin{table*}
\centering
\caption[]{CFHT data points of arcs A,B,C,D upon which orbital solutions were determined. The distance (in km)
is measured from Saturn's center, projected perpendiculary on the equatorial axis of the west ansa. The longitude
is calculated on the basis of orbital solutions given in Table ~\ref{table2}. It is measured from the 
ascending node of Saturn's equatorial plane on the Earth's equator at the standard epoch.
}
\vspace{0.5 cm}
\begin{tabular}{lllll}
\hline 
Object & Decimal day (UTC) & Distance (km) on & Longitude & Filter\\
       & of 1995 November  & equatorial axis  & (degrees) &       \\
\hline 
A &  20.283900 &  125847 & 103.7 & K \\	
A &  20.289005 &  127699 & 105.5 & K \\	
A &  20.293935 &  131403 & 109.5 & K \\	
A &  20.295486 &  132560 & 110.9 & K \\	
A &  20.300613 &  134875 & 114.1 & K \\	
A &  20.301250 &  135570 & 115.2 & K \\	
A &  20.351597 &  135583 & 144.3 & K \\	
\\
B &  20.241563 &  111493 & 92.5  & H \\
B &  20.243727 &  113345 & 93.8  & H \\
B &  20.263044 &  126541 & 104.3 & J \\
B &  20.266250 &  127930 & 105.7 & J \\
B &  20.266944 &  128856 & 106.7 & J \\
B &  20.270891 &  130940 & 108.9 & J \\
B &  20.277755 &  134412 & 113.4 & J \\
B &  20.337269 &  133949 & 146.8 & K \\
B &  20.338669 &  133023 & 148.0 & K \\
B &  20.343287 &  132097 & 149.2 & K \\
\\
C & 20.214919 & 136959 & 117.7 & K \\
C & 20.269444 & 132560 & 148.6 & J \\
C & 20.270359 & 131403 & 150.0 & J \\
C & 20.272118 & 130245 & 151.3 & J \\
C & 20.273333 & 129319 & 152.3 & J \\
C & 20.277755 & 126541 & 155.1 & J \\
C & 20.288356 & 120753 & 160.2 & K \\
C & 20.296134 & 114503 & 164.9 & K \\
C & 20.315706 &  97139 & 175.8 & K \\

\\
D & 20.248773 & 128625 & 153.1 & J \\
D & 20.265012 & 117744 & 162.5 & J \\
D & 20.266250 & 117281 & 162.9 & J \\
D & 20.267894 & 115429 & 164.2 & J \\
D & 20.270891 & 114503 & 164.9 & J \\
D & 20.278275 & 109409 & 168.4 & J \\
\hline
\end{tabular}
\label{table1}
\end{table*}

\begin{table*}
\centering
\caption[]{
Orbital fit for A,B,C,D derived by combining HST and CFHT data. The orbital fit for Pandora is also 
reported as it is detected in both HST and CFHT data (below 80\degr). The standard epoch is 1995 
November 21.5 TDT at Saturn. Longitude is measured from the ascending node of Saturn's equatorial plane on the Earth's equator at the standard epoch.\\
0: Circular orbit.\\
1: Eccentricity and longitude of pericenter of the F ring at the standard epoch are assumed to be 
e=0.0029 and $\omega$=292.5\degr (Synnott et al. \cite{synnott}).\\
2: Eccentricity and longitude of pericenter of Pandora are assumed  to be e=0.0044 and 
$\omega$=274.7\degr.}
\vspace{0.5 cm}
\begin{tabular}{lllll}
\hline 
Object & Semi-major axis & Longitude at epoch &  Residuals  & Mean Motion\\
       &      (km)       &      (degrees)     &    (km)     &  (deg/day) \\
\hline 
A$^{0}$ & 140\,070$\pm$166 & 122.85$\degr\pm$1.3 & 751 & 582.92$\pm$0.96 \\
B$^{0}$ & 140\,083$\pm$108 & 135.81$\degr\pm$0.6 & 617 & 582.80$\pm$0.60 \\ 
C$^{0}$ & 140\,050$\pm$119 & 177.32$\degr\pm$0.8 & 643 & 583.05$\pm$0.68 \\
D$^{0}$ & 140\,046$\pm$141 & 192.47$\degr\pm$0.9 & 691 & 583.07$\pm$0.91 \\ 
 \\
A$^{1}$ & 140\,075$\pm$158 & 122.66$\degr\pm$1.2 & 723 & 582.89$\pm$1.02 \\
B$^{1}$ & 140\,081$\pm$100 & 135.43$\degr\pm$0.6 & 598 & 582.85$\pm$0.61 \\
C$^{1}$ & 140\,017$\pm$109 & 177.66$\degr\pm$0.5 & 611 & 583.23$\pm$0.68 \\
D$^{1}$ & 140\,021$\pm$139 & 192.69$\degr\pm$1.0 & 657 & 583.23$\pm$0.87 \\
Pandora$^{2}$ & 141\,792$\pm$87 & 52.0$\degr\pm$0.3 & 512 & 572.30$\pm$0.53 \\
\hline
\end{tabular}
\label{table2}
\end{table*}

The computation of orbital solutions has been performed using moments J2 to J4 of
Saturn's potential (Campbell \& Anderson \cite{campbell}). The location of objects  A,B,C and D are measured 
on the profiles where the Saturn's scattered light and the 
template of the F ring (the SMP) has been subtracted. The center is determined ``by eye'' by comparing at least 
three successive profiles where a given object appears clearly. Positions are then converted into 
distance from the center of Saturn projected onto the planet's equatorial axis. 
For HST data, profiles in which objects are measured are corrected for the distortion of the PC camera. 
Only the best points were considered with
the longest time baseline, as it is the most important factor for an accurate orbit determination.
The CFHT data points are presented in Table~\ref{table1}, in which the filter as well as the current longitude
 of each object are also reported.\\
Semi-major axis and longitude at epoch are determined by a least square fit of the position of objects 
projected onto the equatorial axis of Saturn. Error bars are obtained by introducing a random noise 
in the input positions, with dispersion equals to the spatial uncertainty 
(0.09\arcsec\,for CFHT and 0.035\arcsec\,for HST). \\
A circular fit was performed first. The results are presented in Table~\ref{table2}. 
They confirm that arcs A to D are located between Pandora and Prometheus 
($\sim 140\,050$ km). Since such elongated structures are expected to be embedded in the F ring, 
we also performed a non-circular fit assuming the eccentricity and longitude of 
pericenter of the F ring, but the residuals did not significantly decrease. \\
If it is assumed that objects A, B, C, D are embedded in the F ring, an estimate of 
its semi-major axis in 1995 is calculated by taking the mean semi-major axis. 
Based on the circular fit we obtain $<a>$=140\,060 $\pm$ 60 km 
($\Leftrightarrow$ n=582.9 $\pm$ 0.4 $\degr/day$), with a physical width of $\pm$20
km around the mean, calculated by taking the standard deviation of the four 
derived semi-major axes. For non-circular fit we obtain $<a_{ex}>$= 140\,050 $\pm$ 60 km
 ($\Leftrightarrow$ n=583.0$\pm$ 0.4 $\degr/day$), with a physical width of $\pm$ 30 km 
around the mean. Combining both data set allows to significantly improve the orbital 
solution of arc B (584$\pm$4 $\degr/day$) as reported in N96. This gives also the first 
determination of the orbits of A,C and D arcs.

\section{Discussion}
\subsection{Comparison with August and November 1995 observations}
Another 10\degr-long arc has been seen by N96 (it is designed here as the ``E arc'', see Fig.~\ref{Fig.2}.c)
on November 21$^{st}$ 1995. Its orbit has
been determined by P2000 using November 17$^{th}$, 18$^{th}$ and 21$^{st}$ HST images. 
Their value of 140\,074 $\pm$ 30 km is consistent ($1\sigma$) with the A,B,C and D arcs location.\\
Some other objects discovered in August 1995, during the Earth 
crossing seemed also to be close to and somewhat below 140\,000 km 
(Roddier et al. \cite{roddier_96a}; N96; Ferrari et. al \cite{ferrari}; P\cite{poulet}). 
However most of them appeared to be point-like, and thus possibly of a different nature than features 
A to E. 
Based on N96 orbital solutions, S6 and S7 are located at  
139\,900 km and 139\,100 km (with error bars of 70 km and 260 km respectively). 
The orbit of S5 is still a matter of debate (P.D. Nicholson, private communication). Its semi-major was 
initially estimated as 139\,860 $\pm$ 130 km by N96 and 140\,050 $\pm$ 100 km by Roddier et 
al.(\cite{roddier_96a}), but two recent studies, combining different data sets, gives uncompatible 
results : P2000 find 140\,208 $\pm$ 50 km and McGhee et al. (submitted to Icarus) report 
a=139690$\pm$ 90 km. A few other objects (S11 to S19) where also discovered by Roddier et al.(\cite{roddier_96a}, 
\cite{roddier_96b}). Their estimated orbits are not accurate enough to give us 
more information on the F ring environment (error bars are $\pm$1000 km and $\pm$2000 km for S8 and S9, 
and $\pm$500 km for objects S11 to S15).\\
\subsection{Comparison with 1980-81 Voyager observations}
Values of $<a>$ and $<a_{ex}>$ are however quite far (3 $\sigma$) from all estimates 
of the F ring radius derived from Voyager images, by about -150 km.
Indeed Synnott et al. (\cite{synnott}) gives a = 140\,185 $\pm$ 30 km. 
A recent study of Voyager images (Ferrari et al. \cite{ferrari}, \cite{ferrari_98}) 
reveals that, at the Voyager epoch, some of the brightest structures of the F ring move on 
different orbits, with a measured dispersion of  95$\pm$15 km ($1\sigma$). 
The measured semi-major axis is 140\,219 $\pm$ 19 km derived 
from Voyager 1 data and 140\,205 $\pm$ 10 km derived from Voyager 2 data. 
All those results are consistent with French and Nicholson who used a 
different approache based on Voyager and  stellar occultation 
data (unpublished work of French \& Nicholson, reported in N96) 
giving a=140\,209 $\pm$ 4 km. 

\subsection{F ring radial structure} 

The width of the F ring in Voyager images is up to 300 km and the ring is formed of 
four eccentric strands (Murray et al.\cite{murray_97}), $F\alpha$, F$\beta$, F$\gamma$ 
and F$\delta$, which radial width is 50 km in average and with respective semi-major 
axes 140\,089 km, 140\,176 km, 140\,219 km and 140\,366 km. The F$\gamma$ strand is 
by far the brightest and may be the ``core'' of the F ring with centimeter-sized bodies 
(Showalter et al. \cite{showalter_92}; Murray et al. \cite{murray_97}), in which 
the brightest features (like clumps or arcs) were detected at the Voyager epoch 
(Ferrari et al. \cite{ferrari}; Showalter \cite{Showalter_97}).
The five arcs seen in  November 1995 (A,B,C,D and E) 
seem to be gathered around 140\,060 km, close to the the faint $F\alpha$ strand. 
Consequently, if the 1995 November arcs belong to a F ring core, this may 
imply a radial restructuring or a spreading of the ring in the meantime.\\

\subsection{Possible implications}
We suggest below, as a direction for future works, that some close encounter with
one of the shepherding satellites may have reshaped the F ring in the last twenty years.
Such an hypothesis may not be excluded since the dynamical evolution 
of this ring is still a puzzle for the scientific community.
The evolution of the F ring  may critically depend on its two shepherding satellites, Pandora and 
Prometheus (see for example: Dermott \cite{dermott}; Showalter \& Burns 
\cite{showalter_82}; Lissauer \& Peale \cite{lissauer}). During the 1995 crossings, 
Prometheus was lagging its expected position by $\sim$ 19\degr\, (N96). It has been 
suggested that an encounter of the F ring with Prometheus may have happened in 
between 1991 and 1994 (Murray \& Giuliatti Winter \cite{murray_96})
due to the precessional variation of the orbits owing to Saturn's oblateness. 
The dynamical consequence of this encounter on the structure of the F ring is not 
currently known. It might result in breaking of the strands and in the creation of 
structures in the inner regions of the F ring. Indeed a massive body is able to 
scatter its environment inside three Hill's radii (Nishida \cite{nishida}; 
Ida \& Makino \cite{ida}) that is about 300 km for Prometheus (assuming a mass 
ratio of $\sim 10^{-9}$ between Prometheus and Saturn). 
Then in case of an instantaneous close approach to the F ring, estimated at $\sim 50$ km 
(Murray \& Giuliatti Winter \cite{murray_96}), Prometheus might be able to 
perturb the ring over a distance of 300 km, i.e. about the full width of the 
F ring. 
Such a model has been recently considered (abstract of Showalter et al. \cite{Showalter_99}; 
P.D. Nicholson, private communicaton), 
and first results show that Prometheus may perturb a portion of the F ring after each close encounter. 
However the resulting radial displacement seems to be about 1 kilometer only. \\
In addition, during the interaction, Prometheus at its  apocenter 
is close to the F ring's pericenter, yielding strong relative velocities of $\sim$ 30 m/s. 
Such  high-velocity encounters may result in catastrophic disruption of bodies initially 
in the F ring. The inner region of the F ring may be consequently populated with 
fragments especially if a belt of kilometer-sized moonlets exists between Prometheus and 
Pandora, as originally suggested by Cuzzi \& Burns (\cite{cuzzi}).   

\section{Conclusion}

  We have reported observations of the crossing of Saturn's ring-plane by the Sun, 
on 1995 November 20$^{th}$, made with the 3.6m CFH Telescope, 
using the UHAO adaptive optics system. Four azimuthally extended structures (A,B,C,D)
have been detected, with central longitude at epoch of 123\degr, 136\degr, 177\degr\, and 
193\degr\ (see Table~\ref{table2}), on 1995 November 21.5 TDT at Saturn. 
These structures have been also seen one day later in HST images (N96). The combination of
 both CFHT and HST data sets provides an accurate estimate of their orbit.
 A circular orbital fit locates them between 140\,020 km and 140\,080 km, 
with error bars of 120 km on average.
The mean circular orbit is 140\,060 $\pm$ 60 km, that is consistent 
with the orbit of another bright arc observed at the same epoch (P2000).
However, this represents a significant discrepancy ($3\ \sigma$) with previous estimates 
of the F ring's radius derived from Voyager images (located at $140\,200$\,km on 
average).
If the five arcs observed in November 1995 belong to a F ring core, this would imply 
a radial restructuring or a spreading of the F ring between 1980 and 1995, whose 
explanation has still to be found. 
In order to understand the F ring's dynamics, better spatial resolution and longer time baselines observations 
are required. The Cassini encounter in 2004 will probably be a major step in the
understanding of the F ring mysteries. 

\begin{acknowledgements}
 The authors thank P.D. Nicholson for his helpful comments, as a referee.

\end{acknowledgements}

\end{document}